\UseRawInputEncoding
\documentclass[journal]{IEEEtran}
\usepackage{doi}
\pdfoutput=1
\usepackage{color}
\usepackage{bm}
\usepackage{amsmath}
\usepackage{esint}
\usepackage{mathrsfs}
\usepackage{amsthm}
\usepackage{graphicx} 
\usepackage{amsfonts} 
\usepackage{subfigure}
\usepackage{booktabs}
\usepackage{float}
\usepackage{multirow}
\usepackage{amssymb}
\usepackage[noadjust]{cite}
\usepackage{calligra}
\usepackage{algorithmicx}
\usepackage[ruled]{algorithm2e}
\usepackage{bbm}

\newtheorem{lemma}{Lemma}
\newtheorem{theorem}{Theorem}
\newtheorem{remark}{Remark}
\newtheorem{assumption}{Assumption}

\newtheorem{definition}{Definition}
\newtheorem{corollary}{Corollary}

\begin{document}

\title{A Control-Recoverable Added-Noise-based Privacy Scheme for LQ Control in Networked Control Systems}

\author{
Xuening Tang, \IEEEmembership{Student Member,~IEEE,}
Xianghui Cao, \IEEEmembership{Senior Member,~IEEE,} and
Wei Xing Zheng, \IEEEmembership{Fellow,~IEEE}
\thanks{Xuening Tang and Xianghui Cao are with the School of Cyber Science and Engineering, Southeast University, Nanjing 210096, China. Xianghui Cao is also with the School of Automation, Southeast University, Nanjing 210096, China. (e-mail: xntang@seu.edu.cn; xhcao@seu.edu.cn).}
\thanks{Wei Xing Zheng is with the School of Computer, Data and Mathematical Sciences, Western Sydney University, Sydney, NSW 2751, Australia
(e-mail: W.Zheng@westernsydney.edu.au).}
\thanks{Corresponding Author: Xianghui Cao (xhcao@seu.edu.cn).}

}

\maketitle

\begin{abstract}
As networked control systems continue to evolve, ensuring the privacy of sensitive data becomes an increasingly pressing concern, especially in situations where the controller is physically separated from the plant. In this paper, we propose a secure control scheme for computing linear quadratic control in a networked control system utilizing two networked controllers, a privacy encoder and a control restorer. Specifically, the encoder generates two state signals blurred with random noise and sends them to the controllers, while the restorer reconstructs the correct control signal. The proposed design effectively preserves the privacy of the control system's state without sacrificing the control performance. We theoretically quantify the privacy-preserving performance in terms of the state estimation error of the controllers and the disclosure probability. Additionally, the proposed privacy-preserving scheme is also proven to satisfy differential privacy. Moreover, we extend the proposed privacy-preserving scheme and evaluation method to cases where collusion between two controllers occurs. Finally, we verify the validity of our proposed scheme through simulations.

\end{abstract}


%
\IEEEpeerreviewmaketitle

\section{Introduction}
\IEEEPARstart {W}{ith} the rapid development of networking and communication technologies, today's communication networks can provide fast and reliable communication between physical devices located in different sites \cite{nowzari2019event}. This remarkable capability has led to the widespread use of communication networks in connecting control components, giving rise to what is known as networked control systems (NCSs). Its most significant feature is to bridge the cyberspace and physical spaces, enabling the flexible configuration of control system components. NCSs have been applied in various fields, including environmental monitoring, industrial automation, robotics, aircraft, automotive, manufacturing, remote diagnosis, as well as remote operation \cite{mahmoud2019networked}.


Despite many benefits of NCSs, several studies have shown that exposing the existing system information to untrusted networked controllers can lead to various security issues, one of the most significant of which is the privacy \cite{wang2022transmission,wang2022decentralized}. The information provided by the local plant might include sensitive data, if analyzed by third-party controllers or eavesdroppers in the communication channel, could potentially compromise the plant's privacy by revealing additional information about its operations. For instance, in connected vehicle systems, drivers are required to share their locations, departure and destination information, which may reveal their home addresses, workplaces, etc. \cite{xu2019edge}. Furthermore, the consequences of privacy leakage in NCSs can be devastating. Consider a scenario where a controller maintains real-time control over the state of a mission Unmanned Aerial Vehicle (UAV), such as flight altitude and speed. If sensitive information like the UAV's trajectory is exposed, then adversaries can easily track and disrupt or directly attack the UAV, resulting in mission failure \cite{zhi2020security}. Therefore, ensuring the privacy security becomes an absolute priority for NCSs.

The main existing privacy-preserving methods in the control domain include homomorphic encryption (HE), algebraic transformation (AT), Added Noise (AN), etc \cite{sultangazin2020symmetries}. 
HE allows computations to be performed on encrypted data without decryption \cite{jia2021blockchain}. In order to address optimization problems on encrypted data, it was recommended in \cite{alexandru2020cloud} to utilize HE for encryption. However, the size of encrypted data is often much larger than the original data, resulting in increased storage requirements \cite{acar2018survey}. AT involves transforming original problems into equivalent ones \cite{weeraddana2013per}. The isomorphism of control systems was presented in \cite{sultangazin2020symmetries} to ensure privacy and trajectory integrity. However, using time-invariant transformation matrices can make it difficult to defend against plaintext attacks \cite{nakano2022known}. Unlike HE and AT, AN is a simpler and easier-to-implement method but may sacrifice control performance for privacy protection. In \cite{9678137}, AN was expanded for utilization in multivariate intelligent network control systems. Additionally, AN was applied to dynamical systems to protect the trajectory data values, as demonstrated in \cite{li2023differentially,yang2022privacy,kawano2021modular}. Therefore, if a privacy-preserving scheme that maintains low complexity and control performance can be designed, then the AN method would be preferred.

Privacy-preserving criteria of AN are typically achieved through methods such as differential privacy (DP) and ($\varepsilon$, $\delta$)-data-privacy. DP protects privacy through a degree of indistinguishability \cite{wang2023tailoring}. DP filtering is discussed in \cite{le2013differentially}, where releasing filtered signals while respecting privacy of user data streams is considered. However, for many control systems, the primary privacy concern is to ensure that adversaries cannot accurately estimate the original data, rather than the indistinguishability \cite{he2018preserving}. In this paper, we propose a novel privacy-preserving scheme based on AN to protect the state trajectory of NCSs. The contribution of this research is threefold.

\begin{enumerate}
\item We propose a novel privacy-preserving scheme utilizing two remote controllers and new noise addition methods. The privacy encoder generates two noise-blurred states and sends them to the controllers, and the restorer recovers the true control with respect to the noise. Thus, the privacy is preserved without sacrificing the control performance. Furthermore, the encoder and the restorer are lightweight with computational complexity as $O(n)$ and $O(m)$, respectively, where $n$ and $m$ are the dimensions of the state and control input, respectively.

\item We evaluate the proposed privacy-preserving scheme and derive the bounds of the state estimation error of the controllers as well as a closed-form expression for the privacy disclosure probability, respectively. The former reflects how far on average the controllers' state estimates are away from the true system's states, while the latter indicates the likelihood that the controllers' state estimation error is within a certain small range. If the added noise of the encoder is Gaussian, then an explicit expression for the disclosure probability is derived. Furthermore, an upper bound is given, which depends only on the dimension of the system state and the covariances of the system and the added noise.

\item Furthermore, we extend our scheme and theoretical analysis to the case that the two controllers collude. We show that, even with colluding controllers, the privacy is still well preserved without sacrificing the linear quadratic (LQ) control performance.

\end{enumerate}


{\it Notations:} $\mathbb R^n$ stands for the $n$-dimensional Euclidean space. $\mathbb{P}(\cdot)$ and ${\mathbb E}(\cdot)$ are the probability and expectation of a random variable, respectively. $\mathbb N(\mu, R)$ represents the Gaussian distribution with mean $\mu$ and covariance $R$. $\top$ is the transpose operator of a vector or matrix. For an $n$-dimensional random variable $X$, $f_{X}(x_{1:n})$ is its probability density function (PDF) and $x_{1:n}=[x_1,x_2,...,x_n]^\top$. For a square matrix $M$, $\mathrm{tr}(M)$ is its trace, and we write $M>0$ and $M\geq 0$ to mean that $M$ is a positive definite and semi-positive definite matrix, respectively. $\Lambda\leq\Upsilon$ means that $\Upsilon-\Lambda$ is a semi-positive definite matrix. $I$ stands for an identity matrix of a proper dimension determined in the context. $\Vert\cdot\Vert$ is the 2-norm of a vector. $diag\{\lambda_1,...,\lambda_n\}$ stands for the diagonal matrix with $\lambda_i$ be the $i$-th diagonal element. $\emptyset$ represents the empty set.

\section{Problem Formulation}
In this section, we introduce the system model and problem statements. Table \ref{tbl:notations} summarizes the definitions of main notations used throughout this paper.

\begin{table}[htbp]
	\centering
	\caption{Important Notations}
	\label{tbl:notations}
	\begin{tabular}{l||l}
		\hline
		Symbol & Definition\\
		\hline
   $\bm{\alpha}(k)$, $\bm{\beta}(k)$ & The noises generated by local plant \\
   $\bm{\tilde{x}}_{\alpha}(k)$, $\bm{\tilde{x}}_{\beta}(k)$ & The blurred state sent to controllers\\
    $\bm{\tilde u}_{\alpha}(k)$, $\bm{\tilde u}_{\beta}(k)$ & The control inputs calculated by controllers \\
    $\bm{\theta}(k)$ & $\bm{\theta}(k)=[\bm{\alpha}^\top(k)\quad \bm{\beta}^\top(k)]^\top$\\
		$R_\alpha$, $R_\beta$, $\bm{R}$ & The covariances of $\bm{\alpha}(k)$, $\bm{\beta}(k)$, $\bm{\theta}(k)$ \\
		$\bm{\hat{x}}(k)$ & Controllers' estimate of $\bm{x}(k)$\\
		$P(k)$ & The error covariance of the \emph{posterior} state estimates\\
	$\delta(k)$ & The disclosure probability of $\bm{x}(k)$\\
		\hline
	\end{tabular}
\end{table}

\subsection{Linear Quadratic (LQ) Formulation}\label{section:2:1}
We consider controlling the following discrete-time linear time-invariant plant:
\begin{equation} \label{system1}
\bm{x}(k+1)=A\bm{x}(k)+B\bm{u}(k)+\bm{w}(k),
\end{equation}
where $k$ is the discrete time step, $\bm{x}(k)\in \mathbb R^n$ is the dynamical state of the plant, $\bm{u}(k)\in \mathbb R^m$ is the control input, and $\bm{w}(k)\in \mathbb R^n$ is a Gaussian white noise with distribution $\bm{w}(k)\sim \mathbb N(0, W)$ and $W>0$. We assume that $(A,B)$ is controllable and $(A,\sqrt{W})$ is stabilizable \cite{zadeh2008linear}.

We consider a finite time horizon of $N$ steps and the control objective is to minimize the following quadratic function:
\begin{equation} \label{cost1}
\begin{aligned}
J(\bm{x},\bm{u})=\frac{1}{N}\mathbb E &\left(\sum^{N-1}_{k=1} \bm{x}^\top(k)Q\bm{x}(k)+\bm{x}^\top(N)Q_N\bm{x}(N)\right.\\
&\left.+\sum^{N-1}_{k=1}\bm{u}^\top(k)U\bm{u}(k)\right),
 \end{aligned}
\end{equation}
where $Q\in\mathbb R^{n\times n}$, $Q_N\in\mathbb R^{n\times n}$ and $U\in\mathbb R^{m\times m}$ are weight matrices with $Q=Q^\top>0$, $Q_N=Q^\top_N>0$ and $U=U^\top>0$. The pair $(A,\sqrt{Q})$ is assumed to be observable \cite{zadeh2008linear}\footnote{This assumption is standard in the LQ control \cite{anderson2007optimal}, which guarantees the existence of solutions to an algebraic Riccati equation.}.

Under \eqref{cost1}, the optimal LQ control input is given below \cite{9691800}:
\begin{equation} \label{u}
	\bm{u}(k)=L(k)\bm{x}(k),
\end{equation}
where 
\begin{align}
	L(k)=&-\left[B^\top S(k+1)B+U\right]^{-1}B^\top S(k+1)A, \label{eq:Lk}\\
	S(k)=&A^\top S(k+1)A+Q-A^\top S(k+1)B \nonumber\\
			 &\times \left[B^\top S(k+1)B+U\right]^{-1}B^\top S(k+1)A,\label{eq:Sk}
\end{align}
and \eqref{eq:Sk} iterates backwards from $S(N)=Q_N$.

\subsection{Problem Statements}\label{section:B}

The above equations in \eqref{u}-\eqref{eq:Sk} imply that, in order to correctly compute the LQ control input, the plant must share with the controller the information: the state $\bm{x}(k)$, the system matrices $A$ and $B$, the parameters $Q,Q_N$ and $U$ in the control objective function, and the time horizon $N$. Among them, the state $\bm{x}(k)$ is considered as the private information of the plant and is not expected to be known by a third party \cite{darup2021encrypted}, while the other system parameters are considered as public information.

Regarding privacy-preserving of $\bm{x}(k)$ by the added noise, the criterion often used in the literature is DP \cite{le2013differentially,hassan2019differential}. However, controllers and eavesdroppers may be able to estimate an approximation of the true state. In this context, the current DP literature does not provide an intuitive or easily explainable way to define adversaries' estimation ability \cite{he2018preserving}. To make up for this deficiency, we introduce the concept of ($\varepsilon$, $\delta$)-data-privacy, which provides a clear and operational metric for privacy protection. This new approach makes the evaluation of privacy risk more explicit and understandable.
\begin{definition}[\cite{he2018preserving}]\label{def:dataprivacy}
An algorithm is said to be of ($\varepsilon$, $\delta$)-data-privacy if for any state $\bm{x}$ and its estimate $\bm{\hat{x}}$ by the algorithm, there holds
\begin{equation}\label{p}
	\mathbb{P} \left(\Vert \bm{x}-\bm{\hat x}\Vert\leq\varepsilon \right)=\delta,
	\end{equation}
where $\delta$ is the disclosure probability and $\varepsilon$ is a small constant.
\end{definition}
\begin{remark}
Definition \ref{def:dataprivacy} uses the $2$-norm to define the distance, which is closely related to our conventional perception of spatial distance. The $2$-norm is also widely used for quantifying error and bias. Given any value of $\varepsilon$, \eqref{p} can calculate the probability that the difference between $\bm{x}$ and $\bm{\hat x}$ is less than $\varepsilon$.
\end{remark}

In this paper, we make the following assumption about the controller.

\begin{assumption}\label{assum:server}
The controller in our scheme is curious but honest \cite{tanaka2017directed}, which means that it may infer the confidential information about the private system state $\bm{x}(k)$ without violating the control law in \eqref{u}.
\end{assumption}

If the local plant sends the raw data of the state $\bm{x}(k)$ to the controller for calculating the LQ control input, then the controller directly knows the trajectory of the private state. Alternatively, if the local plant sends the perturbed state information by means of adding a noise \cite{hassan2019differential}, then the added noise will introduce control biases that, as accumulating over time, will lead to some sacrifice of the control performance \cite{9705530}. Although there are also some other privacy-preserving methods, such as \cite{darup2017towards}, that can effectively prevent a third-party from knowing the system state, they usually impose considerable extra computational overhead on the plant. Thus, we are interested in the following two problems:

\it{Problem 1:} \rm{How to design a lightweight (from the perspective of the local plant) privacy-preserving algorithm to protect the state $\bm{x}(k)$ in \eqref{system1} without sacrificing the LQ control performance \eqref{cost1}?}

\it{Problem 2:} \rm{How to evaluate the performance of the privacy-preserving algorithm when the controller can estimate the
state $\bm{x}(k)$ based on known information?}

From now onwards, we propose a novel privacy-preserving algorithm to address the first problem in Section \ref{yinsi} and evaluate its performance to address the second problem in Section \ref{sec:evaluation}. Section \ref{extend} extends the results to the scenarios that the controllers collude. The relationships between the presented scheme and the existing results in the literature are discussed in more depth in Section \ref{sec:dis}.

\section{Privacy Preserving Algorithm}\label{yinsi}

\begin{figure}[H]
	\centering
\includegraphics[height=5cm,width=8.75cm]{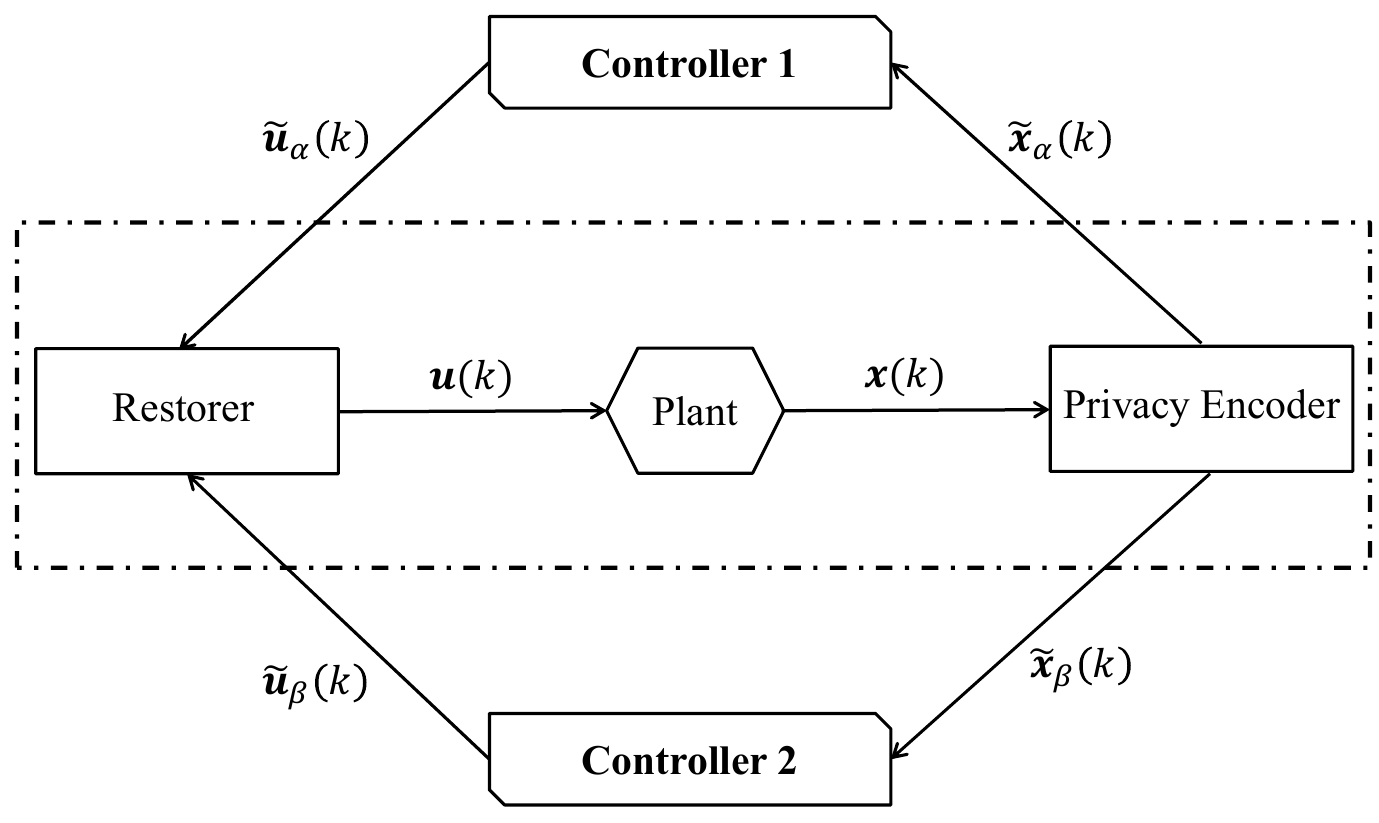}
	\caption{The proposed privacy preserving framework.}
	\label{fig:huitu}
\end{figure}

In numerous existing privacy-preserving mechanisms that use AN strategies, the noise often degrades the performance of the control system, and thus a trade-off between privacy and control performance is often necessary \cite{9705530,kawano2021modular}. To address this pervasive issue, this paper proposes a new approach that utilizes two controllers working together and designs a restorer to eliminate the interference caused by the noise, thereby ensuring privacy preservation while maintaining control performance. The framework of the proposed algorithm is depicted in Fig. \ref{fig:huitu}. In this case, the masking is done by the privacy encoder while the restoration is done by the restorer, both of which are deployed at the local plant. Also, two non-colluding controllers perform the control computation separately (denoted as controller $1$ and controller $2$). The case of colluding controllers is examined in Section \ref{extend}.


At each step $k$, the privacy encoder generates two noises that satisfy the following constraint:
 \begin{align}\label{parallel}
\bm{\alpha}(k)=\rho\bm{\beta}(k),\rho \notin\{0,1\},
\end{align}
where $\bm{\beta}(k)\in\mathbb{R}^n$ is assumed to be identically and independently distributed with any continuous distribution having mean and covariance as $\bm{0}$ and $R_\beta$, respectively. Given a fixed $\rho$, the mean and covariance of $\bm{\alpha}(k)$ are $\bm{0}$ and $R_\alpha=\rho^2R_\beta$. Notice that the choice of the parameter $\rho$ in \eqref{parallel} ensures that $\bm{\alpha}(k)$ and $\bm{\beta}(k)$ are different and that $\bm{\alpha}(k)\neq 0$.

Then, the encoder calculates two blurred states by adding these two noises, i.e.,
 \begin{align}
\bm{\tilde{x}}_{\alpha}(k)&=\bm{x}(k)+\bm{\alpha}(k),\\
\bm{\tilde{x}}_{\beta}(k)&=\bm{x}(k)+\bm{\beta}(k),\label{tildexs2}
\end{align}
and sends them to controller $1$ and controller $2$, respectively. It is assumed that the communication channels between the local side and the controllers are reliable without delay or packet loss. Upon receiving the local packets, the two controllers calculate the LQ control according to \eqref{u}, respectively, i.e.,
\begin{align}
\bm{\tilde u}_{\alpha}(k)=L(k)\bm{\tilde{x}}_{\alpha}(k),\quad \bm{\tilde u}_{\beta}(k)=L(k)\bm{\tilde{x}}_{\beta}(k)\label{ui2}.
\end{align}

After that, the controllers send $\bm{\tilde u}_{\alpha}(k)$ and $\bm{\tilde u}_{\beta}(k)$ back to the local plant, respectively. Upon receiving the controllers' packets, the restorer recovers the correct control input based on \eqref{parallel} and \eqref{ui2}:
\begin{align}
&\frac{\rho}{\rho-1}\bm{\tilde u}_{\beta}(k)-\frac{1}{\rho-1}\bm{\tilde u}_{\alpha}(k)\label{jsuk}\\
=&\frac{\rho}{\rho-1}[\bm{u}(k)+L(k)\bm{\beta}(k)]-\frac{1}{\rho-1}[\bm{u}(k)+L(k)\bm{\alpha}(k)]\nonumber\\
=&\bm{u}(k)\label{uk}.
\end{align}

We solve Problem 1 in Algorithm \ref{algo}: since the true control input is correctly restored, Algorithm \ref{algo} achieves privacy preservation while maintaining the optimal control performance.

\begin{algorithm}\label{algo}
		\caption{Privacy Preserving Algorithm}
		\KwIn{Plant: $\tilde{\bm u}_{\alpha}(k)$, $\tilde{\bm u}_{\beta}(k)$;\\ 
  Controller $1$: $A, B, Q, Q_N, U, N, \tilde {\bm x}_{\alpha}(k)$; \\Controller $2$: $A, B, Q, Q_N, U, N, \tilde{\bm x}_{\beta}(k)$}
		\KwOut{Plant: $\tilde{\bm x}_{\alpha}(k)$, $\tilde{\bm x}_{\beta}(k)$; \\
  Controller $1$: $\tilde{\bm u}_{\alpha}(k)$;\\
Controller $2$: $\tilde{\bm u}_{\beta}(k)$;\\}
		\tcc{Plant:}
		\For{$k = 0,1,2,\ldots,N$}{
			Generate
			 $\bm{\alpha}(k)$ and $\bm{\beta}(k)$ under constraint \eqref{parallel}\;
$\tilde{\bm x}_{\alpha}(k)\leftarrow \bm{x}(k)+\bm{\alpha}(k)$\;
         $\tilde{\bm x}_{\beta}(k)\leftarrow \bm{x}(k)+\bm{\beta}(k)$\;
	        Send $\tilde{\bm x}_{\alpha}(k)$ to controller $1$ and $\tilde {\bm x}_{\beta}(k)$ to controller $2$ respectively\;
	        Restore $\bm{u}(k)$ by \eqref{uk} after receiving $\tilde{\bm u}_{\alpha}(k)$ and $\tilde{\bm u}_{\beta}(k)$ from the controllers\;
		}
		\tcc{Controller $1$:}
		\For{$k = 0,1,2,\ldots,N$}{
		    Compute $S(k)$, $L(k)$ and $\tilde u_{\alpha}(k)$ by \eqref{eq:Lk}, \eqref{eq:Sk} and \eqref{ui2}, respectively\;
			Send $\tilde{\bm u}_{\alpha}(k)$ back to the local plant\;
		}
  \tcc{Controller $2$:}
		\For{$k = 0,1,2,\ldots,N$}{
		    Compute $S(k)$, $L(k)$ and $\tilde u_{\beta}(k)$ by \eqref{eq:Lk}, \eqref{eq:Sk} and \eqref{ui2}, respectively\;
			Send $\tilde{\bm u}_{\beta}(k)$ back to the local plant\;
		}
\end{algorithm}

\begin{remark}
In single-controller NCSs, two approaches can be attempted to eliminate the effects of noise added for privacy protection on the closed-loop system. The first approach involves selecting noise from the null space of $L(k)$, while the second applies multiplicative noise, expressed as $\tilde{\bm{x}}(k)=\varrho(k){\bm{x}}(k)$, where $\varrho(k)$ is a scalar noise variable. However, as $L(k)$ is computed by the controller and the local plant lacks the computational capacity to perform this calculation, predicting $L(k)$ before generating noise becomes impractical, making the first approach infeasible. The second approach, $\tilde{\bm{x}}(k) = \varrho(k){\bm{x}}(k)$, also presents limitations. Since $\varrho(k)$ is a scalar, the direction of ${\bm{x}}(k)$ will either align with or oppose that of $\tilde{\bm{x}}(k)$, thus compromising privacy. Moreover, the effect of $\varrho(k)$ is directly dependent on the magnitude of ${\bm{x}}(k)$. Specifically, $\tilde{\bm{x}}(k) = {\bm{x}}(k) + [\varrho(k) - 1]{\bm{x}}(k)$, indicating that when ${\bm{x}}(k)$ is small, the influence of $\varrho(k)$ is minimal, increasing the risk of information leakage. Conversely, when ${\bm{x}}(k)$ is large, the multiplicative noise can overly distort the state. The dependency on private information $\bm x(k)$ makes it challenging to ensure consistent and effective privacy protection.
\end{remark}

\section{Privacy-preserving Performance Analysis}\label{sec:evaluation}
Algorithm \ref{algo} effectively solves Problem 1 and demonstrates a notable advantage: the design of the restorer ensures that the control performance remains unaffected by the added noise. Consequently, the local plant does not need to make a trade-off between privacy preservation and control performance. To further enhance the level of privacy protection, we can consider adding more noise. However, to evaluate the privacy preservation performance, we consider the worst case that each controller knows the means and covariance matrices $R_\alpha,R_\beta$ of the added noises. However, the precise values and distribution of the noises are unknown to the controllers.

Since the controllers do not collude, they run the same estimation process to estimate $\bm{x}(k)$ independently. Thus, in the following, we shall focus on a generic controller and  drop the subscripts $\alpha$ and $\beta$ in the related notations for ease of exposition. In the rest of this section, we first propose the estimate of the system state for the controller, followed by evaluating the privacy-preserving performance of the proposed algorithm.

\subsection{State Estimation of Controller}

Based on \eqref{system1}, \eqref{u} and \eqref{tildexs2}, the closed-loop control system known to the controller can be rewritten as follows:
\begin{align}
\bm{x}(k+1)&=F(k)\bm{x}(k)+\bm{w}(k),\label{xkk}\\
\bm{\tilde x}(k)&=\bm{x}(k)+\bm{\theta}(k),\label{widetildexk}
\end{align}
where $F(k)=A+BL(k)$, and $\bm{\theta}(k)=\bm{\alpha}(k)$ for controller $1$ and $\bm{\theta}(k)=\bm{\beta}(k)$ for controller $2$.

Besides the parameters $\{A, B, W, Q, Q_N, U, N, R\}$, the information set available to the controller in step $k$ is $\mathcal{I}(k)=\{\bm{\tilde{x}}(1),\bm{\tilde{x}}(2),\ldots,\bm{\tilde{x}}(k)\}$. Denote the \emph{a priori} and \emph{a posterior} estimates of the true state $\bm{x}(k)$ as $\bm{\hat x}(k|k-1)=E(\bm{x}(k)|\mathcal{I}(k-1))$ and $\bm{\hat x}(k)=E(\bm{x}(k)|\mathcal{I}(k))$, respectively. Correspondingly, denote the state estimation errors as $\bm{e}(k|k-1)=\bm{x}(k)-\bm{\hat x}(k|k-1)$ and $\bm{e}(k)=\bm{x}(k)-\bm{\hat x}(k)$, and the state estimation error covariance matrices as
\begin{align}
&P(k|k-1)=\mathbb E(\bm{e}(k|k-1)\bm{e}^\top(k|k-1)\vert\mathcal{I}(k-1)),\\
&P(k)=\mathbb E(\bm{e}(k)\bm{e}^\top(k)\vert\mathcal{I}(k))\label{defpk}.
\end{align}

In view of the linearity of the system model in \eqref{xkk}-\eqref{widetildexk} but without knowing the distributions of the noises, perhaps the best that the controller can do is to run a Kalman filter, known as the linear minimum mean square error (LMMSE) estimator, to estimate the state \cite{yang2022privacy}:
\begin{align}
\bm{\hat{x}}(k)=\bm{\hat {x}}(k|k-1)+G(k)\left[\bm{\tilde x}(k)-\bm{\hat {x}}(k|k-1)\right],\label{eq:KF:x}
\end{align}
where the prediction step is $\bm{\hat {x}}(k|k-1)=F(k-1)\bm{\hat{x}}(k-1)$, and the \emph{priori} error covariance matrix is $P(k|k-1)=F(k-1)P(k-1)F^\top(k-1)+W$. The \emph{posteriori} error covariance matrix is given by
\begin{align}
P(k)=\left[I-G(k)\right]P(k|k-1),\label{eq:KF:sigma}
\end{align}
and $G(k)=P(k|k-1)\left[P(k|k-1)+R\right]^{-1}$.

Based on \eqref{tildexs2}, the optimal estimate of the initial state is $\bm{\hat{x}}(0)=\bm{\tilde x}(0)$ and
the covariance matrix is $P(0)=R$. With the above estimation process, the controller can obtain a rough estimate of the true system state with a bounded estimation error, as described below.

\begin{theorem}\label{theorem:error bounds}
The error covariance matrix $P(k)$ of the state estimate is bounded as follows:
\begin{equation}\label{eq:sigmabounds}
\mathrm{tr}\left((W^{-1}+R^{-1})^{-1}\right) \leq \mathrm{tr}\left(P(k)\right) \leq \mathrm{tr}(R),
\end{equation}
where $P(k)$ is defined and given as in \eqref{defpk} and \eqref{eq:KF:sigma}, respectively.

\end{theorem}

\begin{IEEEproof}  
See Appendix \ref{appendix B}.
\end{IEEEproof}  

From the local plant's point of view, Theorem \ref{theorem:error bounds} tells that, in order to mislead the controller' state estimation to a larger extent, the added noises should have a larger covariance $R$, which is in line with our intuition. It is observed that the state estimation errors of both controller $1$ and controller $2$ exhibit an upward trend as $R_\beta$ increases (note that $R_\alpha$ increases with both $\rho$ and $R_\beta$).


\subsection{Privacy-Preserving Performance Analysis}\label{section:continuous}

In order to derive the disclosure probability of the state estimate $\bm{\hat x}(k)$, we provide the following lemma that establishes the relationship between the estimation error $\bm{e}(k)$ and the noises $\bm{\theta}(k)$ and $\bm{w}(k)$.

\begin{lemma}\label{lemma:svd}
The state estimation error is 
\begin{align}
\bm{e}(k)=C(k)\bm{\gamma}(k) \label{chixi},
\end{align}
where $\bm{\gamma}(k)=[\bm{\theta}^\top(0),\ldots,\bm{\theta}^\top(k),\bm{w}^\top(0),\ldots,\bm{w}^\top(k-1)]^\top\in\mathbb{R}^{(2k+1)n}$, and $C(k)=[C_1(k),\ldots,C_{2k+1}(k)]\in\mathbb{R}^{n\times(2k+1)n}$ with $C_i(k)$ as shown in \eqref{cik} at the top of the next page.
\begin{figure*}
	{\noindent} 
\begin{align}\label{cik}
    \begin{split}
        C_i(k)=\left\{
        \begin{array}{ll}
            -\prod^k_{t=1}[I-G(k+1-t)]F(k-t),&\text{if } i=1,  \\
-\left(\prod^k_{t=i}[I-G(k+i-t)]F(k+i-t-1)\right)G(i-1),&\text{if } i=2,\ldots,k,\\   
-G(k),&\text{if } i=k+1,\\
\left(\prod^{k}_{t=i-k}[I-G(k+i-t)]F(k+i-t-1)\right)[I-G(i-1)],&\text{if } i=k+2,\ldots,2k,\\
I-G(k),&\text{if } i=2k+1.
        \end{array}
    \right.
    \end{split}
\end{align}
\rule[-10pt]{18.07cm}{0.1em}	
\end{figure*}

\end{lemma}
\begin{IEEEproof}  
See Appendix \ref{appendix C}.
\end{IEEEproof}

Since $C(k)$ is not necessarily a full row rank matrix, the dimension of the random variable $\bm{e}(k)$ may be reduced. In this case, to quantify the disclosure probability, we consider 
``compressing" the random variable $\bm{e}(k)$ into a non-singular low-dimensional random variable $\bm{\xi}(k)$. Specifically, by singular value decomposition (SVD), the matrix $C(k)$ is decomposed as $O(k)\Sigma(k)V(k)$
where $O(k)\in\mathbb{R}^{n\times n}$ and $V(k)\in\mathbb{R}^{(2k+1)n\times (2k+1)n}$ are orthogonal matrices, i.e., $O^{\top}(k)O(k)=V^{\top}(k)V(k)=I$. Moreover, $\Sigma(k)$ is denoted as $$\Sigma(k)=\begin{bmatrix}
		 \Sigma_{r(k)}(k) & \bm{0} \\
		\bm{0} & \bm{0} \\
	\end{bmatrix}\in \mathbb{R}^{n\times (2k+1)n},$$
where $\Sigma_{r(k)}(k)=diag\{\lambda_1(k),\lambda_2(k),\ldots,\lambda_{r(k)}(k)\}\in\mathbb{R}^{r(k)\times r(k)}$ with $\{\lambda_i(k),i=1,2,\ldots,r(k)\}$ being the non-zero singular values of $C(k)$ and $r(k)$ being the number of those non-zero singular values, and $\bm{0}$ is the zero matrix of compatible dimensions. Define two new random variables as follows:
\begin{align}
\chi(k)&=V(k)\boldsymbol{\gamma}(k)\in\mathbb{R}^{(2k+1)n},\\
\xi(k)&=\Sigma_{r(k)}(k)\chi_{1:r(k)}(k)\in\mathbb{R}^{r(k)}.
\end{align}

Next, the disclosure probability can be obtained, as shown in the following theorem.

\begin{theorem}\label{theorem:dispro} Let $f_{\bm{\theta}(k)}$ be the PDF of $\bm{\theta}(k)$. Given a small tolerance bound $\varepsilon>0$, the disclosure probability of $\bm{x}(k)$ to the controller is
\begin{equation}\label{th:deltak}
\delta(k)=\int_{\Omega(k)}f_{\bm{\xi}(k)}\left(\bm{\xi}(k)\right)\mathrm{d}\bm{\xi}(k),
\end{equation}
where
\begingroup 
\allowdisplaybreaks 
\begin{align} 
&f_{\bm{\xi}(k)}(\bm{\xi}(k))=f_{\bm{\chi}_{1:r(k)}(k)}( \Sigma_{r(k)}\bm{\xi}(k))\vert{ \Sigma^{-1}_{r(k)}}(k)\vert,\label{pdfxi}\\
&f_{\bm{\chi}_{1:r(k)}(k)}\left(\bm{\chi}_{1:r(k)}(k)\right)\nonumber\\
&\quad=\int_{\mathbb{R}^{(2k+1)n-r(k)}}f_{\bm{\chi}(k)}\left(\bm{\chi}(k)\right)\mathrm{d}\bm{\chi}_{r(k)+1:(2k+1)n}(k),\label{pdfchir}\\
&f_{\bm{\chi}(k)}\left(\bm{\chi}(k)\right)=f_{\bm{\gamma}(k)}\left(V^{-1}(k)\bm{\chi}(k)\right)\vert V^{-1}(k)\vert,\label{pdfchi}\\
&f_{\bm{\gamma}(k)}(\bm{\gamma}(k))=\prod^k_{t=0}f_{\bm{\theta}(t)}(\bm{\gamma}_{tn+1:(t+1)n}(k))\nonumber\\
&\quad\quad\quad\quad\quad\quad\times\prod^{k-1}_{t=0}f_{\bm{w}(t)}\left(\bm{\gamma}_{(t+k+1)n+1:(t+k+2)n}(k)\right),\label{pdfgamma}\\
&f_{\bm{w}(k)}(\bm{w}(k))=\frac{1}{\sqrt{(2\pi)^n\vert W\vert}}\exp\left(-\frac{1}{2}\bm{w}^\top(k) W^{-1}\bm{w}(k) \right),\label{pdfw}\\
&\Omega(k)=\left\{\bm{\xi}(k) \bigg|\sqrt{\sum^{r(k)}_{i=1}\xi^2_i(k)}\leq\varepsilon\right\}\label{omega}.
\end{align}
\endgroup 
\end{theorem}
\begin{IEEEproof}  
See Appendix \ref{appendix D}.
\end{IEEEproof}

In particular, if the added noise $\bm{\theta}(k)$ is a Gaussian noise, then an explicit upper bound of the disclosure probability can be derived as follows.

\begin{theorem}\label{upper} If the AN follows an $n$-dimensional Gaussian distribution, then
\begin{align}
\delta(k)=&\int_{\Omega(k)}\frac{1}{\sqrt{(2\pi)^n\vert P(k)\vert}}\exp\left(-\frac{1}{2}\bm{z}^\top P^{-1}(k)\bm{z}\right)\mathrm{d}\bm{z},\label{gaosidelta}
\end{align}
where the integration area 
\begin{align}
\Omega(k)=\left\{\bm{z}\bigg| \sqrt{\sum^n_{i=1}z_i^2}\leq\varepsilon\right\}, 
\end{align}
$\forall k\in\{1,\ldots,N\}$, and $\delta(k)$ is upper bounded by $\bar\delta$, where
\begin{align}
\bar\delta=\min\left\{\frac{\varepsilon^{n}\sqrt{\vert R^{-1}+W^{-1}\vert}}{2^{\frac{n}{2}}\Gamma(\frac{n}{2}+1)}
,1\right\},
\end{align}
with $\Gamma$ representing the Gamma function
\begin{align}\label{Gamma}
\Gamma(\frac{n}{2}+1)=\left\{
\begin{array}{ll}
(\frac{n}{2})!, & \text{if n is even,} \\
\frac{n!!\sqrt{\pi}}{2^{\frac{n+1}{2}}}, & \text{if n is odd.}
\end{array}\right.
\end{align}
\end{theorem}

\begin{IEEEproof}  
See Appendix \ref{appendix upper}.
\end{IEEEproof}  
Theorem \ref{theorem:dispro} presents a precise approach to calculate the worst-case privacy disclosure probability, accounting for a generic distribution of the noise $\bm{\beta}(k)$ (so for $\bm{\alpha}(k)$). On the other hand, Theorem \ref{upper} establishes the explicit expressions for the disclosure probability and its upper limit when Gaussian noises are introduced. Notably, Theorem \ref{upper} highlights that the upper bound diminishes with increasing noise covariances $R$ and $W$, as well as the system's dimension $n$. This implies that, in systems with higher dimensions and larger noise covariances for both the added noise and plant noise, the controller is harder to access the state's privacy.


\section{The Case When The Controllers Collude}\label{extend}

In this section, with a slight abuse of notation, we continue to use the same notation as in previous sections to refer to the corresponding meanings in the case of colluding controllers. This should not cause confusion, as the context is clear.

\subsection{Privacy-Preserving Algorithm}\label{when ESs collude}

If the two controllers collude by sharing their information received from the local plant, then the noise generation method \eqref{parallel} becomes insecure. This is because if the controllers have access to the means and covariance matrices $R_\alpha,R_\beta$, then they can derive the value of $\rho$ by exploiting the multiplicative relationship between $R_\alpha$ and $R_\beta$ (i.e., \eqref{parallel} implies $\rho=\sqrt[2n]{\vert R_\alpha\vert/\vert R_\beta\vert}$). Additionally, by comparing the observations $\bm{\tilde x}_{\alpha}(k)$ and $\bm{\tilde x}_{\beta}(k)$, the controllers can derive $(\rho-1)\bm{\beta}(k)$. As a result, they can determine the specific noise values and thus obtain $\bm{x}(k)$. To circumvent this, we let $\rho$ be time-varying. Specifically, we randomly and independently select each $\rho(k)$ with equal probability from the set $\mathcal{M}=\{m_1,\ldots,m_M\}$ with $\mathcal{M}\cap\{0,1\}=\emptyset$, i.e., 
\begin{align}\label{alphak}
\mathbb{P} \left(\rho(k)=m_i \right)=\frac{1}{M},\quad \forall m_i\in \mathcal{M}.
\end{align}
Suppose that the mean of $\mathcal{M}$ is $0$, denoted as $\mu_{\rho}=0$, and consequently its variance is $R_{\rho}={\sum^M_{i=1}m^2_i}/{M}$.
With $\rho(k)$ generated as the above, the two noises generated by the privacy encoder turn into
\begin{align}\label{Parallel}
\bm{\alpha}(k)=\rho(k)\bm{\beta}(k).
\end{align}
Since $\rho(k)$ is independent of $\bm{\beta}(k)$, we have $\bm{\alpha}(k)$ with mean $\bm{0}$ and covariance matrix 
\begin{align}
R_\alpha=&\mathbb{E}\left(\rho^2(k)\bm{\beta}(k)\bm{\beta}^\top(k)\right)-\mathbb{E}\left(\rho(k)\bm{\beta}(k)\right)\mathbb{E}^\top\left(\rho(k)\bm{\beta}(k)\right)\nonumber\\
=&R_{\rho}\left[R_\beta+\mathbb{E}(\bm{\beta}(k))\mathbb{E}^\top(\bm{\beta}(k))\right]=R_{\rho}R_{\beta}.\label{Ralpha}
\end{align}
Accordingly, the restorer in \eqref{uk} is changed into
\begin{equation}
\begin{aligned}
\bm{u}(k)=\frac{\rho(k)}{\rho(k)-1}\bm{\tilde u}_{\beta}(k)-\frac{1}{\rho(k)-1}\bm{\tilde u}_{\alpha}(k).
\end{aligned}
\end{equation}

\begin{remark}
Given that $\rho(k)$ is a scalar, the controllers may infer directional information about the noise $\bm \alpha(k)$ and $\bm\beta(k)$. However, the exact magnitude remains unknown, thereby preventing the controllers from determining both the direction and magnitude of the state ${\bm x}(k)$. Specifically, the state vector can be expressed as $\bm x(k)=\tilde{\bm x}_\alpha(k)-{\bm \alpha}(k)=\tilde{\bm x}_\alpha(k)-\Vert\bm \alpha(k)\Vert\hat{\bm \alpha}(k)$, where $\hat{\bm \alpha}(k)$ is a unit vector in the direction of ${\bm \alpha}(k)$, and $\Vert\bm \alpha(k)\Vert$ represents its magnitude. Since $\Vert\bm \alpha(k)\Vert$ is unavailable to the controllers, both the direction and magnitude of ${\bm x}(k)$ cannot be accurately determined. Consequently, this uncertainty effectively blurs the privacy information ${\bm x}(k)$.
\end{remark}

\subsection{Privacy-Preserving Performance Analysis}

With the two controllers colluding, they are able to use both $\bm{\tilde x}_{\alpha}(k)$ and $\bm{\tilde x}_{\beta}(k)$ to estimate $\bm{x}(k)$. Therefore, the combined observation can be written as
\begin{align}
\tilde{\boldsymbol x}(k)=\left[\begin{array}{l}\bm{x}(k)\\
\bm{x}(k)
\end{array}\right]+\left[\begin{array}{l}\bm{\alpha}(k)\\
\bm{\beta}(k)
\end{array}\right]=H\bm{x}(k)+\bm{\theta}(k),\label{ext}
\end{align}
where $H=[I,I]^\top$, $\boldsymbol{\theta}(k)=[\bm{\alpha}(k),\bm{\beta}(k)]^\top$. The mean of $\boldsymbol{\theta}(k)$ are $\bm{0}$ and the covariance matrix is 
\begin{align}
\boldsymbol{R}&=\mathbb{E}(\bm{\theta}
(k)\bm{\theta}^\top
(k))-\mathbb{E}(\bm{\theta}
(k))\mathbb{E}^\top(\bm{\theta}
(k))=\left[\begin{array}{ll}R_\alpha & \bm 0\\
\bm 0 & R_\beta
\end{array}\right],\label{bmR}
\end{align}
where we have used the fact that $\mathbb{E}(\rho(k)\bm{\beta}(k)\bm{\beta}^\top(k))-\mathbb{E}(\bm{\beta}(k))\mathbb{E}(\rho(k)\bm{\beta}^\top(k))=\bm 0$.

Then, the controllers can estimate $\bm{x}(k)$ based on \eqref{xkk} and \eqref{ext}. Like \eqref{eq:KF:x}, the controllers can apply the following LMMSE estimator:
\begin{align}
&\bm{\hat{x}}(k)=\bm{\hat {x}}(k|k-1)+G(k)\left[\bm{\tilde x}(k)-H\bm{\hat {x}}(k|k-1)\right]\label{coleq:KF:x}, \end{align}
where $G(k)=P(k|k-1)H^\top\left[HP(k|k-1)H^\top+\bm{R}\right]^{-1}$ and $P(k)=\left[I-G(k)H\right]P(k|k-1)$. Analogous to Theorem \ref{theorem:error bounds}, in the scenario where the two controllers collude, we have the following result.

\begin{corollary}\label{coltheorem:error bounds}
The error covariance of the state estimate is bounded as follows: 
\begin{align}
&\mathrm{tr}\left(P(k)\right)\geq \mathrm{tr}\left(\left(H^\top \bm{R}^{-1}H+W^{-1}\right)^{-1}\right),\label{eq:sigmabounds1}\\
&\mathrm{tr}\left(P(k)\right)\leq \mathrm{tr}\left((H^\top \bm{R}^{-1}H)^{-1}\right).\label{eq:sigmabounds2}
\end{align}
\end{corollary}
\begin{IEEEproof}  
See Appendix \ref{Appendix colp}.
\end{IEEEproof}  

Combined with \eqref{bmR} and the fact that $H^\top\bm{R}^{-1}H=R^{-1}_\alpha+R^{-1}_\beta$, Corollary \ref{coltheorem:error bounds} indicates that to make the controllers' state estimation more misleading, $\rho$ should have a larger variance $R_\rho$ and $\bm\beta(k)$ should have a larger covariance $R_\beta$.

Similar to Lemma \ref{lemma:svd}, the state estimation error is 
\begin{align}\label{colek}
\bm{e}(k)=C(k)\boldsymbol{\gamma}(k),
\end{align}
where $\boldsymbol\gamma(k)=[\bm{e}^\top(0),\bm{\alpha}^\top(1),\ldots,\bm{\alpha}^\top(k),\bm{\beta}^\top(1),\ldots,\bm{\beta}^\top(k),\\\bm{w}^\top(0),\ldots,\bm{w}^\top(k-1)]^\top\in\mathbb{R}^{(3k+1)n}$ with $\bm{e}(0)=\bm{\alpha}(0)$ or $\bm{e}(0)=\bm{\beta}(0)$. $C(k)=[C_1(k),\ldots,C_{3k+1}(k)]\in\mathbb{R}^{n\times(3k+1)n}$ is shown in \eqref{colcik} at the top of the next page, where $G(k)=\left[G_1(k)\quad G_2(k)\right]\in\mathbb{R}^{n\times 2n}$ with $G_1(k)\in\mathbb{R}^{n\times n}$ and $G_2(k)\in\mathbb{R}^{n\times n}$. Thus, $\bm{e}(k)$ can be rewritten as 
\begin{align}\label{colsvdpsi}
\bm{e}(k)=O(k)\Sigma(k)V(k)\boldsymbol{\gamma}(k), 
\end{align}
where $O(k)\Sigma(k)V(k)$ is an SVD of the matrix $C(k)$. 

\begin{figure*}
	{\noindent} 
\begin{align}\label{colcik}
    \begin{split}
        C_i(k)=\left\{
        \begin{array}{ll}
            -\prod^k_{t=1}[I-G(k+1-t)H]F(k-t),&\text{if }i=1,\\
-\left(\prod^k_{t=i}[I-G(k+i-t)H]F(k+i-t-1)\right)G(i-1),&\text{if }i=2,\ldots,k,\\
-G_1(k),&\text{if }i=k+1,\\
-\left(\prod^{k}_{t=i-k}[I-G(k+i-t)]F(k+i-t-1)\right)G(i-1),&\text{if }i=k+2,\ldots,2k,\\
-G_2(k),&\text{if }i=2k+1,\\
\left(\prod^{k}_{t=i-2k}[I-G(k+i-t)H]F(k+i-t-1)\right)[I-G(i-1)H],&\text{if }i=2k+2,\ldots,3k,\\
I-G(k)H,&\text{if }i=3k+1.\\
        \end{array}
    \right.
    \end{split}
\end{align}
\rule[-10pt]{18.07cm}{0.1em}	
\end{figure*}

Next, we derive a result similar to Theorem \ref{theorem:dispro}, which presents an expression for the disclosure probability when two controllers collude. 

\begin{theorem}\label{coldis} Let $f_{\bm{\beta}(k)}$ be the PDF of $\bm{\beta}(k)$. Given a small tolerance bound $\varepsilon>0$, the disclosure probability of $\bm{x}(k)$ to the controllers is
\begin{equation}
\delta(k)=\int_{\Omega(k)}f_{\bm{\xi}(k)}(\bm{\xi}(k))\mathrm{d}\bm{\xi}(k),
\end{equation}
where the integration area  $\Omega(k)$ is given in \eqref{omega}, $f_{\bm{\xi}(k)}(\bm{\xi}(k))$ is given in \eqref{pdfxi}, \eqref{pdfchir}, \eqref{pdfchi} and  
\begingroup 
\allowdisplaybreaks 
\begin{align} 
&f_{\bm{\gamma}(k)}(\bm{\gamma}(k))=\frac{1}{M}\sum^M_{i=1}\prod^k_{t=0}f_{\bm{\beta}(t)}(\bm{\gamma}_{tn+1:(t+1)n}(k))\nonumber\\
&\quad\times\prod^k_{t=0}\mathcal{D}(\bm{\gamma}_{(t+k+1)n+1:(t+k+2)n}(k)-m_i\bm{\gamma}_{tn+1:(t+1)n}(k))\nonumber\\
&\quad\times\prod^{k-1}_{t=0}f_{\bm{w}(t)}(\bm{\gamma}_{(t+2k+2)n+1:(t+2k+3)n}(k)),\label{colpdfgamma}
\end{align}
\endgroup
where $\mathcal{D}(\cdot)$ represents Dirac delta function and $f_{\bm{w}(t)}$ is given in \eqref{pdfw}.

\end{theorem}
\begin{IEEEproof}  
Following the similar procedure as in \eqref{pdfgamma}, one infers \eqref{colpdfgamma}. Similar to the proof of Theorem \ref{theorem:dispro}, we can obtain $\bm{\xi}(k)$ and $r(k)$ by ``compressing" $\bm{e}(k)$ and $f_{\bm{\xi}(k)}$. Therefore, the disclosure probability can be obtained in a very similar way as in Theorem \ref{theorem:dispro}.
\end{IEEEproof}

Specifically, if $\bm{\beta}(k)$ follows an $n$-dimensional Gaussian distribution, then we can obtain a result similar to Theorem \ref{upper}.

\begin{theorem}\label{colupper} If $\bm{\beta}(k)$ is a Gaussian noise, then an upper bound of $\delta(k)$ is given by
\begin{align}\label{3}
\delta(k)\leq \bar\delta=\min\left\{\frac{\varepsilon^{n}\vert R^{-1}_\alpha+R^{-1}_\beta+W^{-1}\vert\Delta}{2^{\frac{n}{2}}\Gamma(\frac{n}{2}+1)},1\right\},
\end{align}
where $\Delta=\frac{1}{M}\sum^M_{i=1}\sqrt{\frac{\vert R_\beta\vert}{1+m^2_i/R^{2n}_\rho}}$ and $\Gamma$ is shown in \eqref{Gamma}.
\end{theorem}

\begin{IEEEproof}  
See Appendix \ref{Appendix last}.
\end{IEEEproof}  


Theorems \ref{coldis} and \ref{colupper} serve as counterparts to Theorems \ref{theorem:dispro} and \ref{upper}, respectively, in the colluding case. Further, Theorem \ref{colupper} reveals that the upper bound decreases as the system’s dimension $n$, $W$, and $R_\alpha$ increase, while $R_\alpha$ increases with $R_\rho$ and $R_\beta$.

\section{Discussion}\label{sec:dis}
In this paper, AN is chosen as the preferred method for privacy protection, primarily based on several key considerations. HE provides strong privacy protection, but the complexity of this technique may result in significant computational and time overheads, as well as increased data volume and storage costs \cite{marcolla2022survey}. AT requires a trade-off between privacy and computational complexity. A time-invariant transformation matrix may expose the system to plaintext attacks \cite{wang2015secure}, while a time-varying matrix increases the computational cost for local plants. Compared to the first two methods, AN is favoured for its simplicity and ease of implementation. It uses the randomness of the noise to provide effective protection against known plaintext attacks. However, the disadvantage is that the noise can compromise the control performance. In particular, the proposed two-controller scheme is shown to achieve privacy protection without sacrificing control performance. Therefore, AN emerges as the preferred method for addressing the privacy challenges outlined in this paper.

Algorithm \ref{algo} is computationally lightweight. The computational complexity of the encoder is as low as $O(n)$, which is in the same order as the existing AN-based methods \cite{li2023differentially} and \cite{9705530}. However, the computational complexity of the proposed restorer is much lower than many existing methods. For example, in \cite{sultangazin2020symmetries}, the computational complexity of the recovery process is $O(m^3)$ due to matrix inversions and multiplications. In contrast, the computational complexity of $\bm{u}(k)$ in \eqref{jsuk} is $O(m)$.

While the proposed scheme ensures privacy protection without compromising control performance, it may incur other costs, such as computational and communication costs. Compared to two-controller NCSs, which do not recover the true control input, the proposed scheme introduces only minimal computational cost. Specifically, the restoration process in \eqref{jsuk}, as previously analyzed, has a complexity as low as $O(m)$, which is a completely acceptable cost. Compared to single-controller NCSs, the proposed scheme incurs additional hardware costs due to an additional controller utilization. Moreover, it requires the local plant to transmit two fuzzy states $\bm{\tilde{x}}_{\alpha}(k)$ and $\bm{\tilde{x}}_{\beta}(k)$ at each time $k$, potentially increasing communication energy consumption. However, it is important to note that advances in modern communication technologies offer various strategies to mitigate this challenge to some extent. For example, optimizing communication resource allocation can partially reduce additional energy consumption \cite{feng2012survey}. Given the significant advantages of the proposed scheme in maintaining control performance, we consider that the additional hardware and communication costs introduced by the two-controller scheme are acceptable.

The selection of $(\varepsilon,\delta)$-data-privacy as the privacy metric provides a precise quantification of the privacy guarantee offered by local plants against the LMMSE estimator. Moreover, $\delta$ provides a clear operational meaning for the privacy metric, which could be potentially lacking in the DP literature \cite{he2018preserving}. Next, we can show that the proposed scheme can also satisfy DP, simultaneously. Before presenting the results, the adjacency for trajectories are defined as follows.
\begin{theorem}\label{th:dp}
Assume that the privacy parameters $\epsilon>0$ and $\gamma\in(0,1/2)$, and the adjacency parameter $b>0$. If $\bm{\alpha}(k)$ is i.i.d. following $\mathbb{N}(0,R_{\alpha})$ and $\bm{\beta}(k)$ is i.i.d following $\mathbb{N}(0,R_{\beta})$, where $R_{\alpha}=r^2_\alpha I$, $R_{\beta}=r^2_\beta I$ and $r_\alpha,r_\beta\geq\frac{b}{2\epsilon}(K+\sqrt{K^2+2\epsilon})$ with $K=\mathcal{Q}^{-1}(\gamma)$, then the blurred state trajectories $\bm{\tilde x}_{\alpha}=[\bm{\tilde x}^\top_{\alpha}(0),\bm{\tilde x}^\top_{\alpha}(1)\cdots,\bm{\tilde x}^\top_{\alpha}(N)]^\top\in\mathbb{R}^{nN}$ and $\bm{\tilde x}_{\beta}=[\bm{\tilde x}^\top_{\beta}(0),\bm{\tilde x}^\top_{\beta}(1),\cdots,\bm{\tilde x}^\top_{\beta}(N)]^\top\in\mathbb{R}^{nN}$ are $(\epsilon,\sigma)$-DP, where $\bm{\tilde x}_{\alpha}(k)=\bm{\tilde x}(k)+\bm\alpha(k)$ and $\bm{\tilde x}_{\beta}(k)=\bm{\tilde x}(k)+\bm\beta(k)$.
\end{theorem}
\begin{IEEEproof}  
See Appendix \ref{appendix th:dp}.
\end{IEEEproof}  

Notably, Algorithm \ref{algo} effectively eliminates the effect of the added noise, allowing us to select the noise without any specific constraints. This ensures that the chosen noise meets the conditions of Theorem \ref{th:dp}, thus guaranteeing that the proposed scheme satisfies DP. The theoretical support provided by Theorem \ref{th:dp} reinforces our privacy-preserving scheme and further enhances its effectiveness.

\section{Simulation Study} 
In this section, we simulate the proposed privacy-preserving algorithm and evaluate its performance using a dynamical system with parameters (see the reference \cite{kiumarsi2014reinforcement} $A=[-1,2;2.2,1.7]$, $B=[2;1.6]$, $Q=Q_N=diag\{6,6\}$, $U=1$ and $W=[1.87,0.61;0.61,1.34]$, and noise $\bm{\beta}(k)$ as Gaussian $\mathbb{N}(\bm{0},R_\beta)$ with $R_\beta=diag\{3,3\}$. Without controller collusion, setting $\rho=2$ yields $\bm{\alpha}(k)\sim \mathbb{N}(\bm{0},R_\alpha)$ with $R_\alpha=diag\{6,6\}$. The initial state is $\bm{x}(0)=[1.5,0]^\top$, with other initial conditions set to zero. We consider a time horizon $N=60$ and $\varepsilon=0.4$.

\begin{figure}[!ht]
	\centering
	\subfigure[Controller $1$.]{
\includegraphics[height=4.5cm,width=7cm]{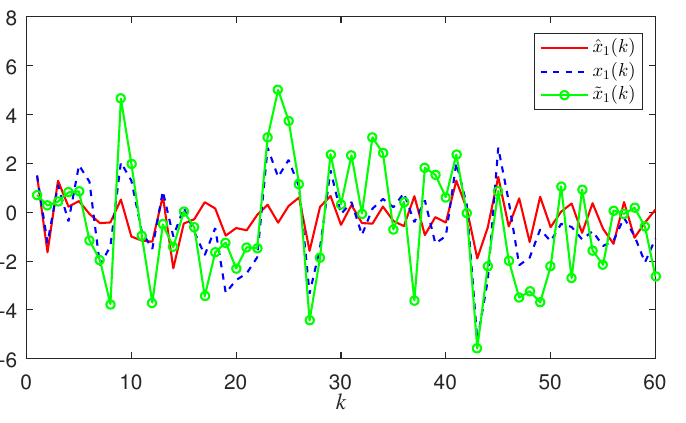}
		\label{fig:a}
	}
	\subfigure[Controller $2$.]{
		\includegraphics[height=4.5cm,width=7cm]{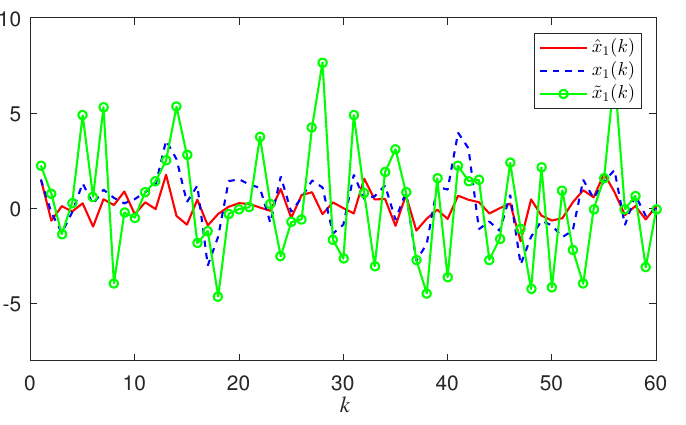}
		\label{fig:b}
	}
 \subfigure[Controller $1$.]{
\includegraphics[height=4.5cm,width=7cm]{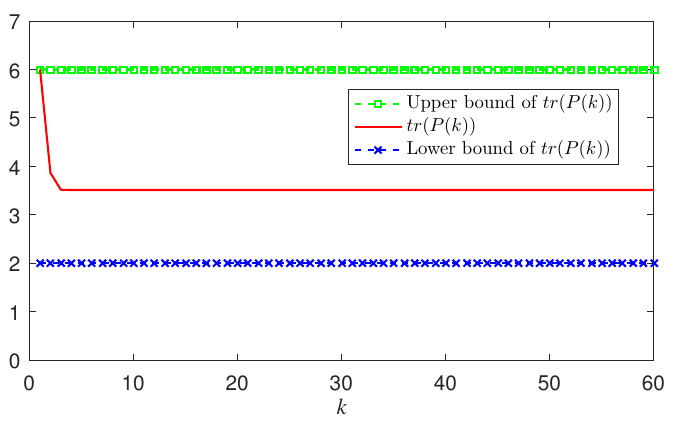}
		\label{fig:c}
	}
	\subfigure[Controller $2$.]{
		\includegraphics[height=4.5cm,width=7cm]{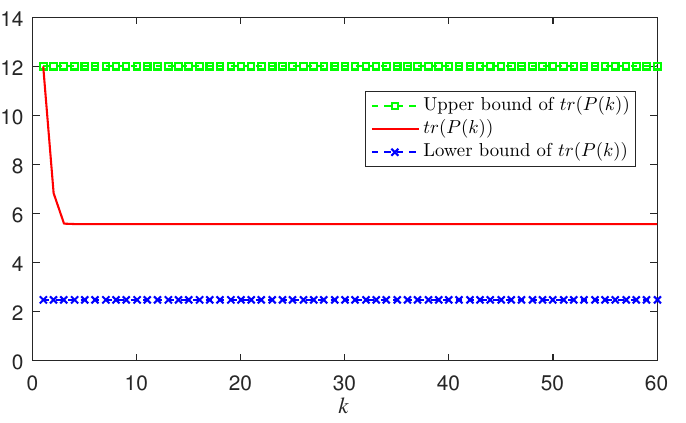}
		\label{fig:d}
	}
	\caption{The trajectories of $x(k)$, $\tilde x(k)$, $\hat x(k)$ and the state estimation performance in the non-collusion case.}
	\label{fig:ac}
\end{figure}
\begin{figure}[!ht]
	\centering
\includegraphics[height=3.85cm,width=8.25cm]{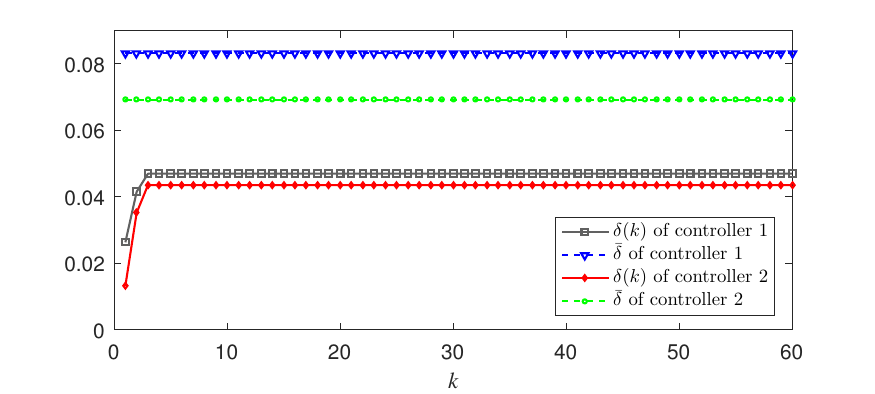}
 \caption{The trajectories of $\delta(k)$ and $\bar\delta$ and in the non-collusion case.}\label{fig:delta}
\end{figure}

Fig. \ref{fig:a} illustrates the true state signal $x_1(k)$, the blurred state $\tilde{x}_{1}(k)$ received by controller $1$ and the corresponding state estimate $\hat x_1(k)$. Similarly, Fig. \ref{fig:b} does the same for controller 2. It can be seen that the blurred state trajectories are significantly different from the real state trajectory. Also, the optimal estimated trajectories still deviate from the true trajectory, and further adjustment of the parameters can lead to better results. Fig.\ref{fig:c} and Fig. \ref{fig:d} displays the error covariance matrix trajectories, which, consistent with Theorem \ref{theorem:error bounds}, stay within theoretical bounds. Fig. \ref{fig:delta} shows the disclosure probabilities $\delta(k)$ and their upper bounds $\bar{\delta}$ for controllers $1$ and $2$, respectively. Both $\delta(k)$ trajectories remain under their respective $\bar{\delta}$, aligning with Theorem \ref{upper}. Theorem \ref{colupper} suggests that with Gaussian noise, the error $e(k)$ also follows a Gaussian distribution, making $\delta(k)$ the integral of $e(k)$ within a certain range. As the error covariance $P(k)$ stabilizes, so does $\delta(k)$, as shown in Figs. \ref{fig:c}, \ref{fig:d} and \ref{fig:delta}. A comparison reveals that controller $1$'s $\mathrm{tr}(P(k))$ is less than that of controller $2$, yet controller $1$'s $\bar\delta$ is higher, indicating that higher noise covariance $R$ enhances privacy.

\begin{figure}[!ht]
	\centering
\includegraphics[height=3.85cm,width=8.25cm]{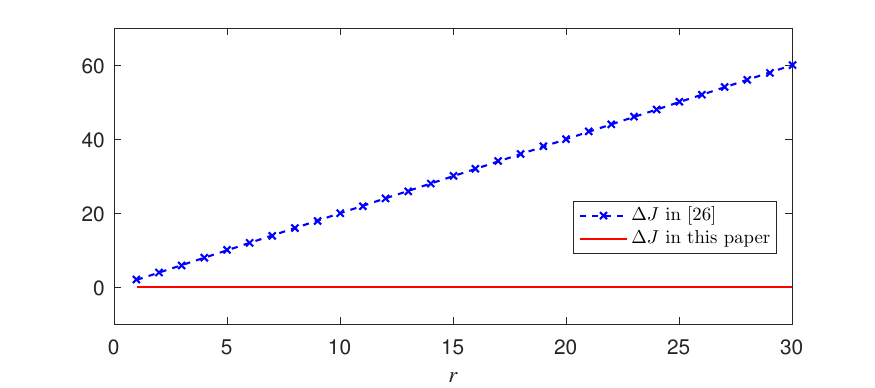}
 \caption{Comparison of the LQ control performance in terms of $\Delta J$ under different schemes in the non-collusion case.}\label{duibi}
\end{figure}

In addition, We compare the proposed method with the state-of-the-art schemes in the references such as \cite{li2023differentially} and \cite{9705530}, and all methods offer equal privacy protection. However, in terms of control performance, our uniquely designed restorer negates the noise's negative impacts, ensuring optimal control. We standardize the noise covariance matrix as $R=rI$. Fig. \ref{duibi} illustrates the $\Delta J$ trajectories for various $R$ values, where $\Delta J=J(x,u)-\bar J(x,u)$, with $\bar J(x,u)$ being the objective in privacy scenarios. Since $\bar J(x,u)$ is identical in \cite{li2023differentially} and \cite{9705530} when all system parameters are the same, we only exhibit $\Delta J$ in our work and \cite{9705530}. The results show that for the scheme in \cite{9705530}, the privacy-related noise leads to the biased objective function, worsening with higher $R$. The proposed restorer maintains control performance, allowing for increased noise and enhanced privacy without compromise, which highlights our scheme's advantage.

\begin{figure}[!ht]
	\centering
\includegraphics[height=3.85cm,width=8.25cm]{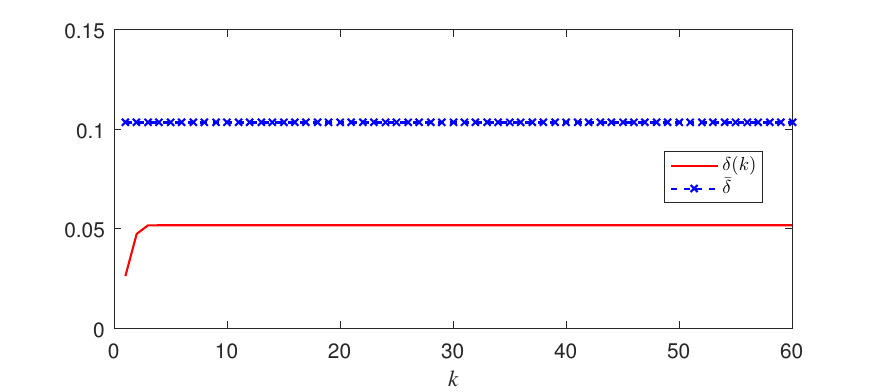}
 \caption{The trajectories of the disclosure probability $\delta(k)$ and its upper bound $\bar\delta$ in the colluding case.}\label{cm}
\end{figure}

Moreover, when the two controllers collude, let $\mathcal{M}=\{\pm 10,\pm 9.9,\ldots,\pm 2.1,\pm2\}$. $\rho(k)$ is selected with equal probability from the set $\mathcal{M}$. The trajectories of the disclosure probability $\delta(k)$ and its upper bound $\bar\delta$ are depicted in Fig. \ref{cm}, which shows a good agreement with Theorem \ref{colupper}.

\section{Conclusion}
This paper has presented a novel privacy-preserving scheme in NCSs. In addition to the low computational overhead, we have formally shown that this approach enables privacy preservation without sacrificing control performance. We have evaluated the proposed privacy-preserving scheme, mainly considering the disclosure probability. Furthermore, we show that the privacy preserving scheme proposed in this paper can also satisfy differential privacy. We have also derived the improved algorithms and evaluations for the privacy-preserving scheme in the two controllers collusion scenario. In the future, we aim to enhance the proposed method to cases when the controllers collude and misbehave and apply the approach of this paper to solve other privacy protection problems in control systems.

\appendix

\subsection{Proof of Theorem \ref{theorem:error bounds}}\label{appendix B}

\begin{IEEEproof} \eqref{eq:KF:sigma} can be rewritten as:
\begin{align}
P(k)&=\left\{I-P(k|k-1)\left[P(k|k-1)+R\right]^{-1}\right\}P(k|k-1) \nonumber \\
&=[P^{-1}(k|k-1)+R^{-1}]^{-1}.\label{ricatipp}
\end{align}
Obviously, we have
\begin{align}\label{pk}
(W^{-1}+R^{-1})^{-1}\leq P(k)\leq R,
\end{align}
where the above inequality applies the fact that $P(k|k-1)\geq W$. Thus, the theorem is proved.
\end{IEEEproof}

\subsection{Proof of Lemma \ref{lemma:svd}}\label{appendix C}
\begin{IEEEproof}  
The optimal estimate of $\bm{x}(k)$ in the LMMSE sense is given by \eqref{eq:KF:x}. Then,
recursively substituting \eqref{eq:KF:x} into $\bm{e}(k)=\bm{x}(k)-\bm{\hat x}(k)$, the following relationship can be obtained:
\begin{align}\label{e}
\bm{e}(k)=&-G(k)\bm{\theta}(k)+[I-G(k)]\bm{w}(k-1)\nonumber\\
&+[I-G(k)]F(k-1)\bm{e}(k-1),
\end{align}
where the initial value of the recursion \eqref{e} is $\bm{e}(0)=-\bm{\theta}(0)$, and $\bm{e}(1)=-G(1)\bm{\theta}(1)+[I+G(1)]\bm{w}(0)-[I-G(1)]F(0)\bm{\theta}(0)$. Thus, when $k\geq 2$, we can get
\begin{align}\label{ecolk}
\bm{e}(k)=&-G(k)\bm{\theta}(k)+[I-G(k)]\bm{w}(k-1)\nonumber \\
&-\sum^{k-1}_{j=1}\left[\prod^{k}_{t=j+1}[I-G(k+j-t)]F(k+j-t-1)\right]\nonumber\\
&\quad\times G(j)\bm{\theta}(j)\nonumber\\
&+\sum^{k-2}_{j=0}\left[\prod^{k}_{t=j+2}[I-G(k+j-t)]F(k+j-t-1)\right]\nonumber\\
&\quad\times[I-G(j+1)]\bm{w}(j)\nonumber\\
&-\prod^k_{t=1}[I-G(k-t+1)]F(k-t)\bm{\theta}(0)\nonumber\\
=&C(k)\bm{\gamma}(k).
\end{align}
Hence, we have completed the proof.
\end{IEEEproof}

\subsection{Proof of Theorem \ref{theorem:dispro}}\label{appendix D}
\begin{IEEEproof} 
Based on Lemma \ref{lemma:svd} and the fact that $\{\bm{\theta}(0),\ldots,\bm{\theta}(k),\bm{w}(0),\ldots,\bm{w}(k-1)\}$ are independent of each other, the PDF of $\bm{\gamma}(k)$ can be expressed as \eqref{pdfgamma}. Since $\bm{w}(t)$ is a Gaussian white noise, $f_{\bm{w}(t)}$ is as shown in \eqref{pdfw}. Thus, \eqref{pdfgamma} is proved.

Since $\bm{\chi}(k)=V(k)\bm{\gamma}(k)$ and $V(k)$ is an orthogonal matrix, there exists a unique inverse transformation
\begin{align}\label{gamma=v-1alpha}
\bm{\gamma}(k)=V^{-1}(k)\bm{\chi}(k),
\end{align}
where $\bm{\gamma}(k)=[\gamma_1(k),\gamma_2(k),\ldots,\gamma_{(2k+1)n}(k)]^\top\in\mathbb{R}^{(2k+1)n}$ and $\bm{\chi}(k)=[\chi_1(k),\ldots,\chi_{(2k+1)n}(k)]^\top\in\mathbb{R}^{(2k+1)n}$. A Jacobian corresponding to this inverse transformation is then calculated as
\begin{align}\label{JA1}
\vert J_{\bm{\gamma},\bm{\chi}}\vert=\left\vert \frac{\partial[\gamma_1(k),\ldots,\gamma_{(2k+1)n}(k)]}{\partial[\chi_1(k),\ldots,\chi_{(2k+1)n}(k)]}\right\vert=\vert  V^{-1}(k)\vert.
\end{align}
Based on the variable transformation method, it follows that
\begin{align}\label{chipdf} 
f_{\bm{\chi}(k)}(\bm{\chi}(k))=f_{\bm{\gamma}(k)}(V^{-1}(k)\bm{\chi}(k))\vert J_{\bm{\gamma},\bm{\chi}}\vert.
\end{align}
Then, \eqref{pdfchi} is proved. Given the PDF $f_{\bm{\chi}(k)}$, the edge PDF $f_{\bm{\chi}_{1:r(k)}(k)}$ is uniquely determined and calculated in \eqref{pdfchir}. Similar to \eqref{gamma=v-1alpha}, since $\bm{\xi}(k)=\Sigma_{r(k)}(k)\bm{\chi}_{1:r(k)}(k)=[\xi_1(k),\ldots,\xi_{r(k)}(k)]^\top$ and $\Sigma_{r(k)}(k)$ is invertible, a Jacobian corresponding to the inverse transformation $\bm{\chi}_{1:r(k)}(k)=\Sigma^{-1}_{r(k)}(k)\bm{\xi}(k)$ is shown below:
\begin{align}
\vert J_{\bm{\xi},\bm{\chi}}\vert=\left\vert \frac{\partial[\chi_1(k),\ldots,\chi_{r(k)}(k)]}{\partial[\xi_1(k),\ldots,\xi_{r(k)}(k)]}\right\vert=\vert \Sigma^{-1}_{r(k)}(k)\vert.
\end{align}
Similar to \eqref{chipdf}, it is derived that
\begin{align}
f_{\bm{\xi}(k)}(\bm{\xi}(k))=f_{\bm{\chi}_{1:r(k)}(k)}( \Sigma_{r(k)}\bm{\xi}(k))\vert J_{\bm{\xi},\bm{\chi}}\vert.
\end{align}
Then, we can obtain the PDF of $\bm{\xi}(k)$ as shown in \eqref{pdfxi}. 

Based on \eqref{chixi} and $C(k)=O(k)\Sigma(k)\bm{\chi}(k)$, we can get
\begin{align}\label{2-norm}
\Vert \bm{e}(k)\Vert=&\Vert O(k)\Sigma(k)\bm{\chi}(k)\Vert=\Vert \Sigma(k)\bm{\chi}(k)\Vert\nonumber\\
=&\Vert\Sigma_{r(k)}(k)\bm{\chi}_{1:r(k)}(k)\Vert =\Vert\bm{\xi}(k)\Vert,
\end{align}
where the second equality holds due to that $O(k)^TO(k)=I$ as $O(k)$ is an orthogonal matrix. According to Definition \ref{def:dataprivacy} and Lemma \ref{lemma:svd}, we have
\begin{align}
\delta(k)&=\mathbb{P}\left(\Vert \bm{e}(k)\Vert\leq\varepsilon \right)=\mathbb{P}\left(\Vert \bm{\xi}(k)\Vert\leq\varepsilon \right)\nonumber\\
&=\mathbb{P}\left(\bm{\xi}(k)\in \Omega(k)\right).
\end{align}
Thus, \eqref{th:deltak} is proved and the proof of the theorem is completed.

\end{IEEEproof}   

\subsection{Proof of Theorem \ref{upper}}\label{appendix upper}

Before proving the theorem, let us first introduce a useful lemma.
\begin{lemma}[\cite{chung2001course}]\label{duoyuan}
A sufficient condition for an $n$-dimensional random variable $\bm{X}=(X_1,X_2,\ldots,X_n)$ obeying an $n$-dimensional Gaussian distribution is that any linear combination
\begin{align}
l_1X_1+l_2X_2+\ldots+l_nX_n
\end{align}
obeys a one-dimensional Gaussian distribution ($l_1,l_2,\ldots,l_n$ are not all zero).
\end{lemma}

\begin{IEEEproof}[Proof of Theorem \ref{upper}] 
Let us first prove that if the added noise $\bm{\theta}(k)$ is an $n$-dimensional Gaussian noise, then $\bm{e}(k)$ follows Gaussian distribution.

Define two new random variables $\bm{\phi}(k)$ and $\bm{\eta}(k)$ as follows:
\begin{align}
\bm{\phi}(k)&=-G(k)\bm{\theta}(k)+[I-G(k)]\bm{w}(k-1),\\
\bm{\eta}(k)&=[I-G(k)]F(k-1)\bm{e}(k-1),
 \end{align}
where $\bm{\phi}(k)=[\phi_1(k),\phi_2(k),\ldots,\phi_n(k)]^\top$ and $\bm{\eta}=[\eta_1(k),\eta_2(k),\ldots,\eta_n(k)]^\top$. Hence, based on \eqref{e}, we have
\begin{align}
\bm{e}(k)=\bm{\phi}(k)+\bm{\eta}(k).
\end{align}
From $G(k)=P(k|k-1)\left[P(k|k-1)+R\right]^{-1}$, we know that the matrices $G(k)$ and $I-G(k)$ are of full rank. If the added noise $\bm{\theta}(k)$ is an $n$-dimensional Gaussian noise, then both $-G(k)\bm{\theta}(k)$ and $[I-G(k)]\bm{w}(k-1)$ follow $n$-dimensional Gaussian distributions, respectively. Since a linear combination of Gaussian distributions is still a Gaussian distribution, the vector $\bm{\phi}(k)$ also obeys an $n$-dimensional Gaussian distribution \cite{chung2001course}. For $\bm{\eta}(k)$, there are two cases: (i) if $F(k-1)$ is of full rank, then $\bm{\eta}(k)$ follows an $n$-dimensional Gaussian distribution, so $\bm{e}(k)$ also follows an $n$-dimensional Gaussian distribution; (ii) if $F(k-1)$ is not of full rank, then $\eta_i(k),i=1,2,\ldots,n$ is a one-dimensional noise or zero. For any nonzero vector $\bm{l}=[l_1,\ldots,l_n]^\top\in \mathbb{R}^n$, it follows that
\begin{align}\label{lll}
&l_1e_1(k)+l_2e_2(k)+\ldots+l_ne_n(k)\nonumber\\
=&l_1\left[\phi_1(k)+\eta_1(k)\right]+\ldots+l_n\left[\phi_n(k)+\eta_n(k)\right].
\end{align}
Since $\bm{\phi}(k)$ is an $n$-dimensional Gaussian noise, $\phi_i(k)$ is a one-dimensional Gaussian noise \cite{chung2001course}. Therefore, it can be seen that the right-hand side of \eqref{lll} is a Gaussian distribution. Based on Lemma \ref{duoyuan}, $\bm{e}(k)$ also follows an $n$-dimensional Gaussian
distribution with $\bm{e}(k)\sim\mathbb{N}(\bm{0},P(k))$, where $P(k)$ is given in \eqref{eq:KF:sigma}. 

Thus, the disclosure probability is given below:
\begin{equation*}
\begin{aligned}
\delta(k)=&\mathbb{P}\left(\Vert \bm{e}(k)\Vert\leq\varepsilon \right)\\
=&\mathbb{P}\left(\bm{e}(k)\in\Omega(k)\right).
\end{aligned}
\end{equation*}
Then, \eqref{gaosidelta} is proved.
Then, it is obvious that the following inequality holds:
\begin{align}\label{underlinedelta}
\delta(k)\leq&\frac{1}{\sqrt{(2\pi)^n\vert P(k)\vert}}\int_{\Omega(k)} \mathrm{d}z\nonumber\\
=&\frac{1}{\sqrt{(2\pi)^n\vert P(k)\vert}}\frac{\pi^{n/2}\varepsilon^{n}}{\Gamma(\frac{n}{2}+1)}.
\end{align}

From \eqref{pk}, we can obtain $\vert P(k)\vert\geq\frac{1}{\vert R^{-1}+W^{-1}\vert}$. Finally, substituting the inequality into \eqref{underlinedelta} completes the proof.\end{IEEEproof}


\subsection{Proof of Corollary \ref{coltheorem:error bounds}}
\begin{IEEEproof}\label{Appendix colp} 
Similar to \eqref{ricatipp}, we get
\begin{align}\label{colPk}
P(k)=\left[H^\top \bm{R}^{-1}H+P^{-1}(k|k-1) \right]^{-1}.
\end{align}
Then, we have
\begin{align}\label{colpk}
\left(H^\top \bm{R}^{-1}H+W^{-1}\right)^{-1}\leq P(k)\leq (H^\top \bm{R}^{-1}H)^{-1},
\end{align}
thus completing the proof.
\end{IEEEproof}

\subsection{Proof of Theorem \ref{colupper}}\label{Appendix last}

Before proving the theorem, let us first introduce a useful lemma.
\begin{lemma}\label{ZD}(\cite{bernstein2009matrix}, Corollary 8.4.15) If both $\Lambda$ and $\Upsilon$ are positive semi-definite matrices of order $n\times n$, then
\begin{equation}
\vert\Lambda+\Upsilon\vert\geq \vert\Lambda\vert+\vert\Upsilon\vert.
\end{equation}
\end{lemma}
\begin{IEEEproof} [Proof of Theorem \ref{colupper}] Similar to \eqref{e}, we have
\begin{align}\label{cole}
\bm{e}(k)=&-G(k)\bm{\theta}(k)+[I-G(k)H]\bm{w}(k-1)\nonumber\\
&+[I-G(k)H]F(k-1)\bm{e}(k-1)\nonumber\\
=&-G_1(k)\bm{\alpha}(k)-G_2(k)\bm{\beta}(k)+[I-G(k)H]\bm{w}(k-1)\nonumber\\
&+[I-G(k)H]F(k-1)\bm{e}(k-1).
\end{align}
Then, from \eqref{bmR}, it can be obtained that
\begin{align}\label{colgk}
G(k)&=P(k)H^\top\bm{R}^{-1}=\left[\begin{array}{ll}P(k)R^{-1}_\alpha&P(k)R^{-1}_\beta
\end{array}\right]\nonumber\\
&=\left[\begin{array}{ll}G_1(k)&G_2(k)
\end{array}\right].
\end{align}
According to \eqref{colPk} and \eqref{colgk}, we can attain 
\begin{align}
\vert G_1(k)\vert&=\frac{1}{\vert R_\alpha\vert}\frac{1}{\vert R^{-1}_\alpha+R^{-1}_\beta+P^{-1}(k|k-1)\vert}\nonumber\\
&\geq\frac{1}{\vert R_\alpha\vert}\frac{1}{\vert R^{-1}_\alpha+R^{-1}_\beta+W^{-1}\vert},\label{g1}
\\
\vert G_2(k)\vert&=\frac{1}{\vert R_\beta\vert}\frac{1}{\vert R^{-1}_\alpha+R^{-1}_\beta+P^{-1}(k|k-1) \vert}\nonumber\\
&\geq\frac{1}{\vert R_\beta\vert}\frac{1}{\vert R^{-1}_\alpha+R^{-1}_\beta+W^{-1}\vert},\label{g2}
\end{align}
where we have applied the fact that $P(k|k-1)\geq W$.

Obviously, both $G_1(k)$ and $G_2(k)$ are full-rank matrices. Then, under the assumption that $\bm{\alpha}(k)$ and $\bm{\beta}(k)$ are Gaussian distributions, following a similar argument as in the proof of Theorem \ref{upper}, it can be deduced that the estimation error $\bm{e}(k)$ in \eqref{cole} also follows a Gaussian distribution.

In the collusion case, since $\bm{\alpha}(k)$ is no longer Gaussian even if $\bm{\beta}(k)$ is Gaussian, the Gaussianity of the estimation error $\bm{e}(k)$ is broken. Therefore, in order to examine the distribution of $\bm{e}(k)$, we first consider the PDF of $\bm{e}(k)$ conditioned on a given sequence of $\boldsymbol{\rho}(k)=(\rho(1),\rho(2),\ldots,\rho(k))$, and then derive the overall PDF of $\bm{e}(k)$ by accounting for the distribution of $\boldsymbol{\rho}(k)$. To this end, consider any given $\boldsymbol{\rho}(k)$, say $\boldsymbol{\rho}(k)=\boldsymbol{m}(k)=(m_{i(1)},m_{i(2)},\ldots,m_{i(k)})$, where $\{m_{i(1)},\ldots,m_{i(k)}\}\subseteq \mathcal{M}$. We have
\begin{equation}
\begin{aligned}
\bm{\alpha}(k)\mid_{\boldsymbol{\rho}(k)=\boldsymbol{m}(k)}=m_{i(k)}\bm{\beta}(k).
\end{aligned}
\end{equation}
In this case, $\bm{\beta}(k)\sim\mathbb{N}(\bm{0},R_\beta)$ implies $\bm{\alpha}(k)\mid_{\boldsymbol{\rho}(k)=\boldsymbol{m}(k)}\sim\mathbb{N}(\bm{0},m^2_{i(k)}R_\beta)$.

Then, based on the above analysis and \eqref{cole}, we can get $\bm{e}(k)\mid_{\boldsymbol{\rho}(k)=\boldsymbol{m}(k)}\sim\mathbb{N}(\bm{0},P_{\bm{e}(k)\mid{\boldsymbol{\rho}(k)=\boldsymbol{m}(k)}}(k))$, where
\begin{align}\label{Ycol}
&P_{\bm{e}(k)\mid{\boldsymbol{\rho}(k)=\boldsymbol{m}(k)}}(k)\nonumber\\
=&G_1(k)m^2_{i(k)}R_\beta G^\top_1(k)+G_2(k)R_\beta G^\top_2(k)\nonumber\\
&+[I-G(k)H]W[I-G(k)H]^\top+[I-G(k)H]F(k-1)\nonumber\\
&\times P_{\bm{e}(k-1)\mid\boldsymbol{\rho}(k-1)=\boldsymbol{m}(k-1)}(k)\nonumber\\
&\times F(k-1)^\top[I-G(k)H]^\top,
\end{align}
with the initial value $P_{\bm{e}(0)\mid \rho(0)=m_{i(0)}}(0)$ being either $R_\alpha$ or $R_\beta$.

Combining \eqref{alphak} and \eqref{Parallel} yields
\begin{align}\label{fecol}
f_{\bm{e}(k)}(\bm{e}(k))=&\overbrace{\mathop{\sum^M_{i=1}P(m_{i(1)}=m_i)\cdots\sum^M_{i=1}P(m_{i(k)}=m_i)}}^{k}\nonumber\\
&f_{\bm{e}(k)\mid\boldsymbol{\rho}(k)}(\bm{e}(k)\mid\boldsymbol{\rho}(k)=\boldsymbol{m}(k)),
\end{align}
where $f_{\bm{e}(k)\mid\boldsymbol{\rho}(k)}(e(k)\mid\boldsymbol{\rho}(k)=\boldsymbol{m}(k))$ represents the PDF of $\bm{e}(k)\mid_{\boldsymbol{\rho}(k)=\boldsymbol{m}(k)}$. Using Lemma \ref{ZD}, \eqref{g1}, \eqref{g2} and \eqref{Ycol}, we have
\begin{align}
&\vert P_{\bm{e}(k)\mid{\boldsymbol{\rho}(k)=\boldsymbol{m}(k)}}(k)\vert\nonumber\\
\geq&\vert G_1(k)\vert\vert m^2_{i(k)}R_\beta \vert\vert G_1(k)\vert+\vert G_2(k)\vert\vert R_\beta\vert\vert G_2(k)\vert\nonumber\\
\geq&\frac{m^2_{i(k)}\vert R_\beta\vert^2+\vert R_\alpha\vert^2}{\vert R_\alpha\vert^2\vert R^{-1}_\alpha+R^{-1}_\beta+W^{-1}\vert^2\vert R_\beta\vert}.\label{YY}
\end{align}
According to \eqref{YY}, we can get
\begin{align}\label{f}
&f_{\bm{e}(k)\mid\boldsymbol{\rho}(k)}(\bm{e}(k)\mid\boldsymbol{\rho}(k)=\boldsymbol{m}(k))\nonumber\\
=&\frac{1}{\sqrt{(2\pi)^n\vert P_{\bm{e}(k)\mid{\boldsymbol{\rho}(k)=\boldsymbol{m}(k)}}(k)\vert}}\nonumber\\
&\times \exp\left[-\frac{1}{2}\bm{e}(k)\vert P_{\bm{e}(k)\mid{\boldsymbol{\rho}(k)=\boldsymbol{m}(k)}}(k)\vert^{-1}\bm{e}^\top(k)\right]\nonumber\\
\leq&\frac{1}{\sqrt{(2\pi)^n\vert P_{\bm{e}(k)\mid{\boldsymbol{\rho}(k)=\boldsymbol{m}(k)}}(k)\vert}}\nonumber\\
\leq &\frac{\sqrt{\vert R_\beta\vert}\vert R^{-1}_\alpha+R^{-1}_\beta+W^{-1}\vert\vert R_\alpha\vert}{\sqrt{(2\pi)^n(m^2_{i(k)}\vert R_\beta\vert^2+\vert R_\alpha\vert^2)}}.
\end{align}

By using Definition \ref{def:dataprivacy}, \eqref{fecol} and \eqref{f}, we can obtain
\begin{align}
\delta(k)=&\mathbb{P}\left(\Vert \bm{e}(k)\Vert\leq\varepsilon \right)=\mathbb{P}\left(\bm{e}(k)\in\Omega(k)\right)\nonumber\\
=&\int_{\Omega(k)}f_{\bm{e}(k)}(z)\mathrm{d}z\nonumber\\
\leq&\int_{\Omega(k)}\sum^M_{i=1}P(m_{i(k)}=m_i)\nonumber\\
&\frac{\sqrt{\vert R_\beta\vert}\vert R^{-1}_\alpha+R^{-1}_\beta+W^{-1}\vert\vert R_\alpha\vert}{\sqrt{(2\pi)^n(m^2_{i(k)}\vert R_\beta\vert^2+\vert R_\alpha\vert^2)}}\mathrm{d}z\nonumber\\
=&\frac{1}{M}\sum^M_{i=1}\frac{\sqrt{\vert R_\beta\vert}\vert R^{-1}_\alpha+R^{-1}_\beta+W^{-1}\vert\vert R_\alpha\vert\pi^{n/2}\varepsilon^{n}}{\sqrt{(2\pi)^n(m^2_i\vert R_\beta\vert^2+\vert R_\alpha\vert^2)}\Gamma(\frac{n}{2}+1)}\nonumber\\
=&\frac{1}{M}\sum^M_{i=1}\frac{\varepsilon^{n}\vert R^{-1}_\alpha+R^{-1}_\beta+W^{-1}\vert}{2^{n/2}\Gamma(\frac{n}{2}+1)}\sqrt{\frac{\vert R_\alpha\vert^2\vert R_\beta\vert}{m^2_i\vert R_\beta\vert^2+\vert R_\alpha\vert^2}}.
\end{align}
Using the fact that $R_\alpha=R_\rho R_\beta$, we therefore complete the proof.
\end{IEEEproof}

\subsection{Proof of Theorem \ref{th:dp}}\label{appendix th:dp}
\begin{IEEEproof}
Let $\bm{x}=[\bm{x}^\top(0),\cdots,\bm{x}^\top(N)]^\top\in\mathbb{R}^{nN}$, $\bm{x}'=[\bm{x}'^\top(0),\cdots,\bm{x}'^\top(N)]^\top\in\mathbb{R}^{nN}$ be two adjacent trajectories, and denote $\bm{v}=\bm{x}-\bm{x}'$. We use the notation $\Vert\cdot\Vert$ for the 2-norm. For $\mathbb{S}\in\mathbb{R}^{nN}$, we have
\begin{align}\label{dp}
&\mathbb{P}(\bm{x}+\bm{\alpha}\in \mathbb{S})\nonumber\\
=&\frac{1}{(2\pi r^2_\alpha)^{\frac{nN}{2}}}\int_{\mathbb{R}^{nN}}\textrm{1}_S(\bm{x}+\bm{\alpha})\exp(\frac{\Vert\bm{\alpha}\Vert^2}{-2r^2_\alpha})\mathrm{d}\bm{\alpha}\nonumber\\
=&\frac{1}{(2\pi r^2_\alpha)^{\frac{nN}{2}}}\int_{\mathbb{R}^{nN}}\textrm{1}_S(\bm{z})\nonumber\exp(\frac{\Vert\bm{z}-\bm x\Vert^2}{-2r^2_\alpha})\mathrm{d}\bm{z}\nonumber\\
=&\frac{1}{(2\pi r^2_\alpha)^{\frac{nN}{2}}}\int_{\mathbb{S}}\exp(\frac{\Vert \bm{z}-\bm{x}'\Vert^2}{-2r^2_\alpha})\nonumber\\
&\times\exp(\frac{2(\bm{z}-\bm{x}')^\top \bm{v}-\Vert \bm{v}\Vert^2}{2r^2_\alpha})\mathrm{d}\bm{z}\nonumber\\
=&\exp^\epsilon\mathbb{P}(\bm{x}'+\bm{\alpha}\in \mathbb{S})\nonumber\\
&+\frac{1}{(2\pi r^2_\alpha)^{n/2}}\int_{\mathbb{S}}\exp(\frac{\Vert \bm{z}-\bm{x} \Vert^2}{-2r^2_\alpha})\textrm{1}_{\{2(\bm{z}-\bm{x}')^\top \bm{v}\geq\Vert \bm{v}\Vert^2+2\epsilon r^2_\alpha\}}\mathrm{d}\bm{z},
\end{align}
where $\bm{\alpha}=[\bm{\alpha}^\top(0),\bm{\alpha}^\top(1)\cdots,\bm{\alpha}^\top(N)]^\top\in\mathbb{R}^{nN}$ and $1_{\{\cdot\}}$ denotes the indicator function. With the change of variables $\bm{y}=\frac{\bm{u}-\bm{x}}{r_\alpha}$ in the integral, we can rewrite it as 
\begin{align}
\mathbb{P}\left(\bm{y}^\top\bm{v}\geq\epsilon r_\alpha-\frac{\Vert \bm{v}\Vert^2}{2 r_\alpha}\right),
\end{align}
with $\bm{y}\sim\mathbb{N}(0,I)$. In particular, $\bm{y}^\top\bm{v}\sim\mathbb{N}(0,\Vert \bm{v}\Vert^2)$,  hence is equal to $\Vert \bm{v}\Vert z$ in distribution, with $z\sim \mathbb{N}(0,1)$. We are then led to set $r_\alpha$ sufficiently large so that \begin{align}\mathbb{P}\left(z\geq\frac{\epsilon r_\alpha}{\Vert \bm{v}\Vert}-\frac{\Vert \bm{v}\Vert}{2 r_\alpha}\right)\leq\gamma,
\end{align}
i.e., 
\begin{align}
\mathcal{Q}(\frac{\epsilon r_\alpha}{\Vert \bm{v}\Vert}-\frac{\Vert \bm{v}\Vert}{2 r_\alpha})\leq \gamma.
\end{align}
Then, according to the definition of $(\epsilon,\sigma)$-differential privacy \cite{le2013differentially}, the result then follows by straightforward calculation. The proof that the trajectory $\bm{\tilde x}_{\beta}$ is also $(\epsilon,\sigma)$-DP is similar to that of $\bm{\tilde x}_{\alpha}$. Thus, we have completed the proof.\end{IEEEproof}

\bibliographystyle{IEEEtran}
\bibliography{manuscript}

\end{document}